\documentclass[epj]{svjour}

\usepackage{graphicx}
\usepackage{fancyhdr}
\def\makeheadbox{{%  remove EPJC header
 }}
\def\mytitle{My title} 
\def\myauthors{My name}  
\def\mytype{My type of session}
\def\mysession{My session}
\def\mytitle{SUSY sensitivity in final states with leptons, jets, and missing transverse energy} 
\def\myauthors{Massimiliano Chiorboli}  
\def\mytype{Contributed Talk}
\def\mysession{Colliders - SUSY Phenomenology}
\newcommand{\chionez}{\ensuremath{\tilde{\chi}^{0}_{1}}} \newcommand{\chitwoz}{\ensuremath{\tilde{\chi}^{0}_{2}}}

\newcommand{\sLepR}{\ensuremath{\tilde{l}_{R}}}

\newcommand{\GeVcc}{\ensuremath{\mathrm{GeV}/c^2}}

\begin{document}
\title{SUSY sensitivity in final states with leptons, jets, and missing transverse energy}
\author{Massimiliano Chiorboli \\
On behalf of the ATLAS and CMS Collaborations
\thanks{\emph{Email:} massimiliano.chiorboli@ct.infn.it}% 
}                     % Do not remove
\institute{Universit\`a di Catania and INFN, Sezione di Catania}
\date{}
\abstract{
Final states with leptons, jets and missing transverse energy are a promising way to detect Supersymmetry at the Large Hadron Collider.
Results of simulation studies performed by the CMS and ATLAS collaborations are presented. The discovery of new physics is shown
to be possible within the first fb$^{-1}$ of collected integrated luminosity. 
 % end of PACS codes
} %end of abstract
\maketitle
\section{Introduction}
\label{intro}
Supersymmetry~\cite{susy} (SUSY) is considered to be one of the most elegant extensions of the Standard Model (SM). 
It is based on a fermion-boson symmetry, built associating
Superpartners to the known particles:
sleptons and squarks are the bosons associated to leptons and quarks, gluinos are the fermions associated to the gluons, 
and the superpartners associated to gauge bosons and to the Higgs scalar mix
into charginos ($\chi_{1,2}^{\pm}$) and neutralinos ($\chi_{1,2,3,4}^0$). The ($\chi_{1}^0$) is the Lightest Supersymmetric Particle (LSP), weakly interacting
and stable in many models. The MSSM has a large number of free parameters~\cite{martin}, and constrained models are often used in simulation studies.
Minimal Supergravity~\cite{msugra} (mSUGRA) has 
only five independent parameters, whose values are fixed at the Grand Unification (GUT) scale: the common gaugino mass m$_{1/2}$, the common scalar mass m$_0$, 
the common trilinear coupling A$_0$ , the ratio of the vacuum expectation values 
of the two Higgs doublets tan$\beta$  and the sign of the Higgsino mixing parameter sign($\mu$).

If SUSY exists, it is expected to manifest itself with a rich phenomenology at the forthcoming Large Hadron Collider (LHC):
squarks and gluinos produced in proton-proton collisions at a center-of-mass energy of 14 TeV yield long decay chains, 
with production of high-p$_{T}$ jets,
leptons, and missing energy due to the Lightest Supersymmetric Particle escaping the detector.
Final states with jets, leptons and missing transverse energy (MET), are hence promising channels for the detection of Supersymmetric particles with
the ATLAS~\cite{atlas} and CMS~\cite{cms} detectors.

\section{Analysis strategies}
\label{strategy}
The ATLAS and CMS Collaborations have recently performed simulation studies to evaluate the reach in discovering Supersymmetry. 
Focus has been given to possible discoveries in the early stages of data taking, within 1~fb$^{-1}$ of integrated luminosity, 
and to model independent studies, considering final states rather than specific channels. 
Both Collaborations performed detailed analyses, using
a full simulation of the detector at benchmark points of the mSUGRA parameter space chosen in order to cover a broad range of possible experimental signatures.
In some cases the effect of misalignments and miscalibrations of the detectors have also been evaluated.
Many possible systematic uncertainties are taken into account, the most important one being
the jet energy scale, assumed to be known at the level of 7\% for 1~fb~$^{-1}$ and 2\% for
10~fb$^{-1}$.
The selections are optimized, and the strategy defined, at a few mSUGRA benchmark points; 
after that, in order to evaluate the discovery reach, a fast simulation software is used to perform a scan of the mSUGRA parameter space.
Details of these studies can be found in~\cite{cms_ptdr_II} and~\cite{atlas_ccs}.
Thanks to the excellent performances of the ATLAS and CMS detectors, 
final states including leptons (electrons and muons) allow a clean and efficient triggering and a stronger background suppression with respect 
to final states with jets and missing transverse energy only~\cite{tytgat}.
Channels with taus in the final states have been studied by both
Collaborations, but are not included in this report.
Typical selection cuts for these events are:
\begin{itemize}
\item $\ge$ n(=1,2,3,4) jets, p$_{T} > $~O(100)~GeV/c;
\item $\ge$ n(=1,2) leptons, p$_{T} > $~O(10)~GeV/c;
\item E$^{T}_{miss} > $~O(100)~GeV
\end{itemize}
The main backgrounds considered for the analyses are $t\overline{t}$, W + jets, Z+jets and QCD multijet processes. Signals are usually produced
using PYTHIA~\cite{pythia} interfaced with ISAJET~\cite{isajet}, while backgrounds are generated with PYTHIA or with 
Matrix Element generators like ALPGEN~\cite{alpgen} or MadGraph~\cite{madgraph}.

\section{Final states with a single lepton, jets and missing transverse energy}
\label{1lep_jet_met} 
\begin{figure}
% For example, with the option graphicx use
\includegraphics[width=0.45\textwidth,angle=0]{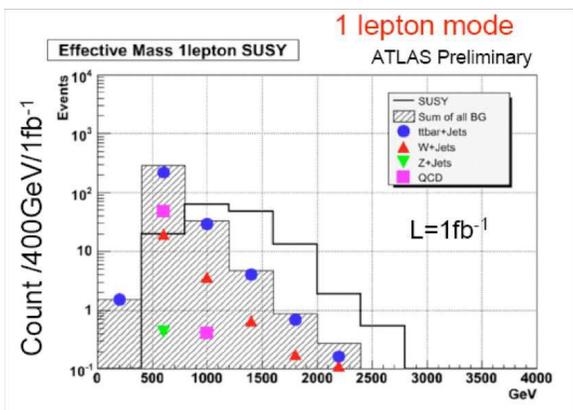}
\caption{ATLAS: effective mass distribution after lepton and jet cuts are applied for the single lepton final state analysis}
\label{fig:atlas_1lepton}       % Give a unique label
\end{figure}

In Supersymmetric events, leptons arise mainly 
from charginos and neutralinos, other possible sources being tops and taus produced in the decay chain.
Events are triggered by a single isolated electron or muon; off-line selection requires
three jets with p$_{T} > $ 50~GeV/c and  E$^{T}_{miss} > $~130~GeV, besides the lepton.
In the CMS analysis~\cite{cms_1lepton} the software package Garcon~\cite{garcon} is used to optimize the cuts
by trying $\sim$O(10$^{50}$) cut set permutations for millions of input events in times of the order of hours.
The fully simulated and reconstructed LM1 mSUGRA point 
(defined by m$_0$ = 60~GeV/c$^2$, m$_{1/2}$ = 250~GeV/c$^2$, tan $\beta$ = 10, A$_0$ = 0, $\mu > $ 0)~\cite{benchmark}
is taken as benchmark for selection optimization and study of systematic effects.
The analysis is hence repeated on fully simulated samples of six different mSUGRA points.
With an integrated luminosity of 10~fb$^{-1}$, a signal to background ratio
larger than 25 is reached in five points, including the effect of the systematic uncertainties.
The large signal to background ratio is confirmed by
the ATLAS analysis, which uses a variable called effective mass, defined as:
\begin{equation}
M_{eff} = E^{T}_{miss} + \Sigma_{jets} p_{T}^{jet}.
\end{equation}
Fig.~\ref{fig:atlas_1lepton} shows the effective mass distribution for signal and background events after the cuts on leptons and jets.

\section{Final states with same sign leptons, jets and missing transverse energy}
\begin{figure}
% For example, with the option graphicx use
\includegraphics[width=0.45\textwidth,angle=0]{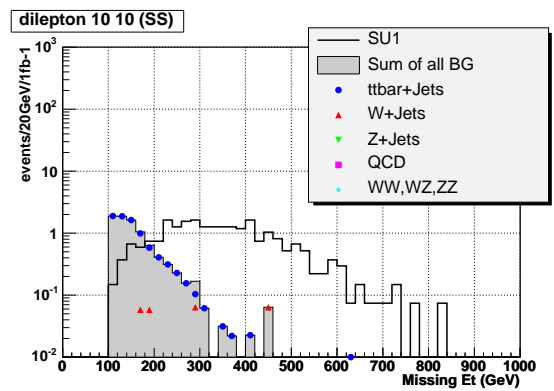}
\caption{ATLAS: missing transverse energy distribution for events selected requiring two same sign leptons.}
\label{fig:atlas_same_sign}       % Give a unique label
\end{figure}

Same-sign leptons can result from several signal processes.
Since gluinos are Majorana particles, they are expected to decay with equal probability into a positive or a negative lepton, thus
50\% of dilepton events coming from gluino pair production have same-sign leptons.
Squark production is another important source 
of same sign dileptons, since the squark charge tends to be determined by the valence quarks in the proton-proton collisions. 
Associated production of superpartners, e.g. gluino-squark, chargino-squark etc., also results in same sign dilepton events.
The CMS analysis focuses on the clean experimental signature of the dimuon event topology, which has in addition the advantage of an efficient and 
well-understood trigger even in the early stages of the LHC data taking. 
The off-line selection requires two same sign leptons in addition to jets and missing transverse energy.
Cuts are optimized on the LM1 mSUGRA point, and the analysis is repeated on fully simulated and reconstructed samples generated
at 10 different mSUGRA points.  
Statistical significances larger than 16 are achieved in 7 out of the 10 mSUGRA points considered for the detailed studies, 
including the effect of the systematic uncertainties.
Fig.~\ref{fig:atlas_same_sign} shows the distribution of the missing transverse energy in two different mSUGRA points 
for events produced with a fast simulation software for the ATLAS analysis.

\section{Final states with opposite sign leptons, jets and missing transverse energy}
\begin{figure}
% For example, with the option graphicx use
\includegraphics[width=0.45\textwidth,angle=0]{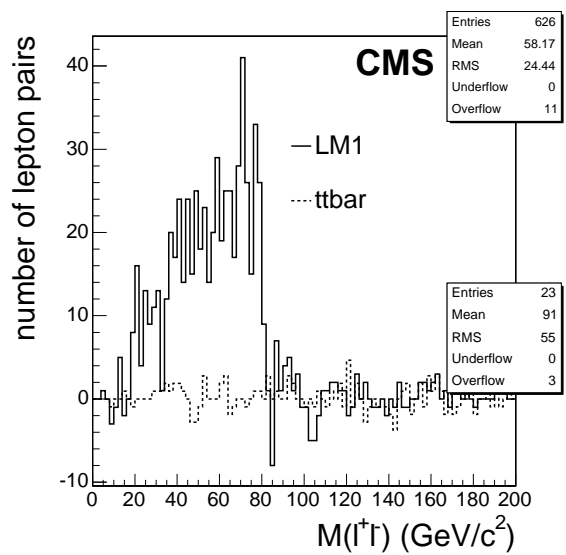}
\caption{CMS: invariant mass distribution of same flavour opposite sign lepton pairs after the subtraction of the 
different flavour opposite sign lepton pair contribution. SUSY signal and $t\overline{t}$ contribution are shown.}
\label{fig:cms_endpoint}       % Give a unique label
\end{figure}

Final states with same-flavour opposite-sign (SFOS) dileptons, originating from the decays $\chitwoz\to\chionez l^{+}l^{-}$ and 
$\chitwoz\to\sLepR l\to \chionez l^{+}l^{-}$, with $l=e,\mu$,
provide a clean signature of SUSY with isolated leptons, high p$_T$ jets and missing transverse energy.
In addition, the dilepton invariant mass distribution of these decays is expected to have a triangular shape
with a sharp upper edge, which makes the signature striking.
The trigger is based on a single isolated electron or muon.
In the CMS analysis~\cite{cms_ofos} a missing energy greater than 200~GeV and two jets
with p$_{T}$ greater than 100 and 60 GeV/c are required to reduce the Standard Model backgrounds. 
In SUSY events the presence of two SFOS leptons can also be due to processes different from $\chitwoz\to\sLepR l\to \chionez l^{+}l^{-}$ decay. 
If the two leptons are independent of each other, one expects equal amounts of SFOS leptons and different flavour 
opposite sign (DFOS) leptons: the background SFOS contribution can hence be removed by subtracting the DFOS events.
Fig.~\ref{fig:cms_endpoint} shows the SFOS lepton pair invariant mass distribution after the DFOS contribution 
has been subtracted; also shown is the contribution from the $t\overline{t}$ process, which is the main Standard Model background.
The endpoint in the distribution is related to the mass of the particles involved in the decay chain: its measurement can therefore give a first information
about the mass scale of the supersymmetric particles produced in the collision. At the benchmark point LM1, the fit to 
the flavour subtracted distribution with a convolution of a triangle and a Gaussian function gives
\begin{equation}
M_{ll}^{max} = 80.42 \pm 0.48 \, \mathrm{GeV}/c^2
\end{equation}
for an integrated luminosity of 1~fb$^{-1}$, where the uncertainty quoted is only statistical. The 
expected value of the endpoint, computed from the simulated masses, is 81.04~\GeVcc.
A statistical significance of 5 sigma is estimated to be achievable with an integrated luminosity of 17~pb$^{-1}$,
including the effect of systematic.
In addition the misalignment of the tracker and muon system
of the CMS detector foreseen for the first months of data taking has been considered. It does not affect dramatically the detection of SUSY events,
lowering the efficiency of dielectron and dimuon selection by 10\% and 30\% respectively, and gives an uncertainty of 1 GeV/c$^{2}$ on the measurement of the
endpoint value.

\section{Searches with Z}
\begin{figure}
% For example, with the option graphicx use
\begin{center}
\includegraphics[width=0.4\textwidth,angle=90]{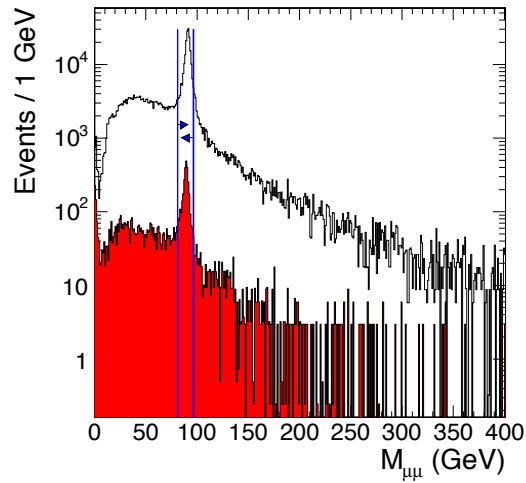}
\end{center}
\caption{CMS: dilepton invariant mass distribution for supersymmetric and standard model events for the analysis searching Z bosons produced by susperymmetric
decay chains.}
\label{fig:cms_z}       % Give a unique label
\end{figure}

SUSY processes leading to final states with Z can be detected using the Z decays into leptons.
The analysis has been performed by the CMS Collaboration at the test point LM4 
(m$_0$ = 210~GeV/c$^2$, m$_{1/2}$ = 285~GeV/c$^2$, tan $\beta$ = 10, A$_0$ = 0, $\mu > $ 0), in which the decay
$\chitwoz\to Z + \chi_1^0$ has a branching ratio close to 100\%.
Events are triggered requiring two electrons or muons. At off-line level,
the invariant mass of the two leptons is required to be in a window around the Z mass~\cite{cms_z}.
In order to reduce the Standard Model background, E$_T^{miss}$ is required to be greater than 230~GeV, and the 
angle between the two leptons is required to be less than 2.65~rad.  Fig.~\ref{fig:cms_z} shows the dilepton invariant
mass distribution for supersymmetric and Standard Model events after the cut on the missing transverse energy; the selected region around the Z peak is also shown.
Five sigma excess can be reached with 100~pb$^{-1}$ of integrated luminosity. Systematic
uncertainties are also taken into account, the most important ones being the lepton p$_T$ resolution, giving a 3\% uncertainty in the number of background events, 
and the E$_T^{miss}$ energy scale uncertainty, estimated to give a 20\% uncertainty in the number of background events.

\section{Events containing top quarks}
In a large region of the mSUGRA parameter space, the supersymmetric partner of the top quark is the lightest squark.
An evidence of its existence can hence be a clear signature for Supersymmetry.
The CMS analysis is optimized at the LM1 point, where the stop decays according to $\tilde{t} \to t \chi_2^0 \to t \sLepR l \to  t l^{+}l^{-} \chionez$.
Events having a single isolated lepton, E$_T^{miss}$ greater than 150~GeV, and four jets with p$_T$ greater than 30~GeV/c, one of which has to be b-tagged, are selected; 
A kinematic  fit is applied to the jets to find the best combination matching the top mass~\cite{cms_top}. 
One of the main systematics is the b-tagging efficiency, whose effect 
to the number of background events
is estimated to be at the level of 8\%. Taking into account 
the systematic uncertainties, an excess of five sigma significance can be found with about 200~pb$^{-1}$.

\section{Parameter space scan}
\begin{figure}
\begin{center}
\includegraphics[width=0.4\textwidth,angle=0]{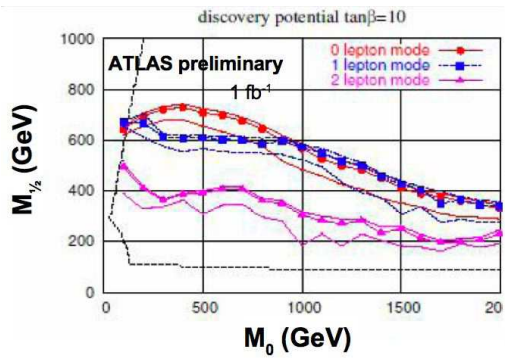}
\end{center}
\caption{ATLAS discovery contours in final states with no leptons, one lepton and two leptons for an integrated luminosity of 1~fb$^{-1}$.}
\label{fig:atlas_curves}       % Give a unique label
\end{figure}

\begin{figure}
\begin{center}
\includegraphics[width=0.4\textwidth,angle=0]{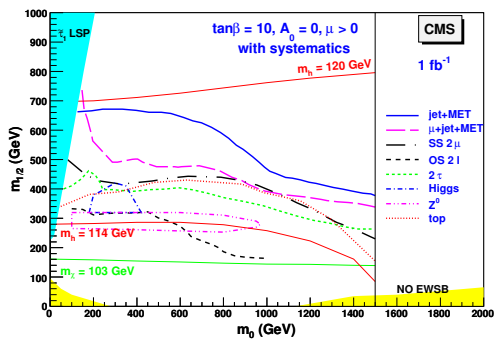}
\end{center}
\caption{CMS discovery contours in various final states with leptons for an integrated luminosity of 1~fb$^{-1}$.}
\label{fig:cms_curves}       % Give a unique label
\end{figure}

The ATLAS and CMS Collaborations have performed a scan of the mSUGRA parameter space using a fast simulation of their detectors. The resulting discovery curves
for an integrated luminosity of 1~fb$^{-1}$ are shown in Fig.~\ref{fig:atlas_curves} for ATLAS and Fig.~\ref{fig:cms_curves} for CMS.
Fig.~\ref{fig:atlas_curves} does not take into account the systematic uncertainties; their inclusion 
is expected to lower the reach in m$_{1/2}$ by about 50~GeV/c$^2$. Therefore, ATLAS is expected to be sensitive to about 
m$_{1/2}$ = 700~GeV/c$^2$ in the one lepton final state.
Fig.~\ref{fig:cms_curves} includes the systematic uncertainties. 
The final state with one lepton yields the best results among the ones with leptons in the final state,
with a reach comparable to that of ATLAS, while the other signatures can help in confirming the discovery. 

\section{Conclusions}
Final states with leptons, jets and missing transverse energy could lead to the discovery of Supersymmetric events in the very early phase of LHC running. 
Many final states have been considered by the ATLAS and CMS collaborations, bringing a redundancy
of possible discoveries. 
The presence of leptons in the final state has the double advantage of an excellent trigger efficiency and more robust analyses with 
large signal to background ratios, providing the possibility to discover new physics with very low integrated luminosities, 
of the order of few tens of pb$^{-1}$.
These discovery capabilities are extrapolations of significances  calculated with statistical samples 1 or 10~fb$^{-1}$. 
A good understanding of the detector and of systematic uncertainties in the very early phases of data taking will therefore be crucial
for the success of the experiments.

\section*{Acknowledgments}

The author wishes to thank D. Costanzo for providing useful material for ATLAS. Special thanks also to G. Rolandi, A. Tricomi and L. De Nardo 
for their help with the manuscript.

%
% BibTeX users please use
% \bibliographystyle{}
% \bibliography{}

\end{document}